\documentclass[aps,pre,twocolumn,groupedaddress,longbibliography]{revtex4}

\usepackage{graphicx}
\usepackage{bm}
\usepackage{amsmath}

\begin{document}

\title{Minimization principle for shear alignment of liquid crystals}
\author{Xingzhou Tang}
\author{Jonathan V. Selinger}
\affiliation{Department of Physics, Advanced Materials and Liquid Crystal Institute, Kent State University, Kent, Ohio 44242, USA}

\date{December 26, 2019}

\begin{abstract}
If a static perturbation is applied to a liquid crystal, the director configuration changes to minimize the free energy.  If a shear flow is applied to a liquid crystal, one might ask:  Does the director configuration change to minimize any effective potential?  To address that question, we derive the Leslie-Ericksen equations for dissipative dynamics, and determine whether they can be expressed as relaxation toward a minimum.  The answer may be yes or no, depending on the number of degrees of freedom.  Using theory and simulations, we consider two specific examples, reverse tilt domains under simple shear flow and dowser configurations under plane Poiseuille flow, and demonstrate that each example shows relaxation toward the minimum of an effective potential.
\end{abstract}

\maketitle

\section{Introduction}

In experiments on nematic liquid crystals, the director field can be aligned by many different physical mechanisms, including electric and magnetic fields, surface anchoring, and shear flow.  All of these mechanisms are useful for technological applications.  In liquid-crystal theory, there is an important difference between the theoretical approaches that are used to describe these alignment mechanisms.  If the alignment is induced by a static perturbation, it is modeled by adding appropriate terms to the free energy function, and then minimizing the free energy over all possible director fields.  By contrast, if the alignment is induced by shear flow, it is modeled by solving hydrodynamic equations~\cite{Ericksen1960,Ericksen1961,Leslie1966,Leslie1968,Beris1994}.  When we compare these approaches, it is natural to ask:  Is it possible to describe shear alignment by any type of minimization principle?  In other words, does the nematic director field evolve to minimize any effective potential?  If so, then we could develop an intuitive picture of shear alignment as motion toward some optimal state, by analogy with alignment by an applied field or other static perturbation.

To our knowledge, this kind of question has been considered at least three times in the liquid-crystal literature.  First, Olmsted and Goldbart developed a theory for the nonequilibrium isotropic-nematic transition under shear flow~\cite{Olmsted1990,Olmsted1992}.  This theory shows that shear flow has the same effect as an aligning field on the isotropic-nematic transition:  Weak shear flow raises the transition temperature, and sufficiently strong shear flow induces an isotropic-nematic critical point.  Because the theoretical approach involves solving hydrodynamic equations rather than minimizing any potential, it is easy for the theory to determine the limit of stability for each phase, but it is difficult for the theory to find the first-order transition temperature.  The first-order transition cannot be identified by searching for the minimum of any function; rather, it must be calculated by modeling the motion of the isotropic-nematic interface.

Later, Doi developed a general theory for the dynamics of soft matter, based on an Onsager-type variational approach~\cite{Doi2011,Doi2013}.  This theory is based on a single scalar function, called the Rayleighian, which combines the dissipation function with the generalized velocities and with derivatives of the energy function.  This approach is a variational theory, because it derives the equations of motion by setting certain derivatives of the Rayleighian equation to zero.  However, it is not exactly a minimization principle, because the system does not evolve toward the global minimum of any function.  This distinction will be discussed in Sec.~II below.

Most recently, Emer{\v{s}}i{\v{c}} \emph{et al}.\ developed theory and simulations to model the behavior of ``dowser'' and ``bowser'' domains in nematic liquid crystals confined in a narrow cell under Poiseuille flow~\cite{Emersic2019}.  Part of their work involves defining an effective potential or effective free energy for the dowser state, which includes the effects of flow as well as the elastic free energy.  In this case, the director field does evolve toward the minimum of the effective potential, and one can see that Poiseuille flow causes the dowser state to become more favorable than the bowser.

The purpose of this article is to introduce a unified theoretical formalism to address this issue.  We want to determine when liquid-crystal dynamics under shear flow can be described by an effective potential, such that the system moves toward the minimum of that potential.

In Sec.~II, we begin with a simple analogy in classical mechanics, in which we consider a particle moving in the wind.  In one dimension (1D), it is straightforward to define an effective potential for this particle, which includes the dissipative effects of the wind.  By comparison, in two dimensions (2D), this effective potential is only defined if the wind velocity field has zero curl, or if the motion is restricted to a 1D track on the 2D plane.

In Sec.~III, we apply this concept to a uniform liquid crystal under simple shear flow.  We show how the effects of shear flow can be represented by an effective potential.  In particular, the shape of the effective potential determines whether the director will tumble or align at the Leslie angle~\cite{Leslie1966,Leslie1968}.

In Sec.~IV, we extend the theory to a nonuniform liquid crystal with no defects.  In particular, we describe the dynamics of reverse tilt domains under simple shear flow, and demonstrate that the domain walls move in order to minimize an effective potential.

In Sec.~V, we generalize the theory to a nonuniform liquid crystal with defects.  In particular, we consider the disclination at the boundary between the dowser and bowser states in a narrow cell, and show that this disclination moves to minimize an effective potential.  For this problem, our results are consistent with the theory of Emer{\v{s}}i{\v{c}} \emph{et al}.~\cite{Emersic2019}; we show that their effective potential fits into the general framework presented here.

Finally, in Sec.~VI, we provide a general discussion of these examples.  We argue that the concept of minimizing an effective potential is a useful theoretical tool for describing liquid crystals under shear flow.  In the problems where it applies, the effective potential particularly helps to develop intuition for the effects of shear alignment.

\section{Classical mechanics analogy}

To introduce the concept of an effective potential for dissipative forces, we present a simple analogy in classical mechanics.  We first demonstrate this concept in 1D, and then show its limitations in higher dimensions.

\subsection{One dimension}

Consider the dynamics of a classical particle in the wind.  This particle experiences a conservative force from its potential energy, as well as a dissipative force from air drag against the wind.  The equation of motion can be written as
\begin{equation}
m\ddot{x}=F_\text{conservative}+F_\text{dissipative}
=-\frac{\partial U}{\partial x}-\frac{\partial D}{\partial\dot{x}},
\end{equation}
where $x(t)$ is the particle position, $m$ is the mass, $U$ is the potential energy, and $D$ is the Rayleigh dissipation function.  For overdamped motion, this equation simplifies to
\begin{equation}
0=-\frac{\partial U}{\partial x}-\frac{\partial D}{\partial\dot{x}}.
\label{overdamped}
\end{equation}
In steady state, we have $\dot{x}=0$, and hence the equation simplifies further to
\begin{equation}
0=-\frac{\partial U}{\partial x}-\left[\frac{\partial D}{\partial\dot{x}}\right]_{\dot{x}=0}.
\label{steadystate}
\end{equation}

For a simple model of the conservative force, suppose the potential energy is $U(x)=\frac{1}{2}k (x-x_0)^2$, so that $F_\text{conservative}=-k(x-x_0)$.  For a simple model of the dissipative force, suppose the wind velocity field is $v_\text{wind}(x)$.  In that case, the Rayleigh dissipation function is $D=\frac{1}{2}\gamma[\dot{x}-v_\text{wind}(x)]^2$, where $\gamma$ is the drag coefficient, and $F_\text{dissipative}=-\gamma[\dot{x}-v_\text{wind}(x)]$.  Hence, the overdamped equation of motion becomes
\begin{equation}
0=-k(x-x_0)-\gamma[\dot{x}-v_\text{wind}(x)],
\end{equation}
and the steady-state equation is
\begin{equation}
0=-k(x-x_0)+\gamma v_\text{wind}(x).
\end{equation}
Thus, we see that the wind shifts the steady-state position of the particle from $x_0$ to $x_0+(\gamma/k)v_\text{wind}(x)$.

To express the steady-state solution as a minimization principle, we would like to rewrite the steady-state equation (\ref{steadystate}) in the form 
\begin{equation}
0=-\frac{\partial U_\text{eff}}{\partial x}.
\end{equation}
Hence, we must define the effective potential $U_\text{eff}(x)$ as
\begin{equation}
U_\text{eff}(x)=U(x)+\int dx \left[\frac{\partial D}{\partial\dot{x}}\right]_{\dot{x}=0}.
\end{equation}
For the simple example above, this construction gives
\begin{equation}
U_\text{eff}(x)=\frac{1}{2}k (x-x_0)^2 -\int dx \gamma v_\text{wind}(x).
\end{equation}
If the wind velocity is uniform, this effective potential is just $U_\text{eff}(x)=\frac{1}{2}k (x-x_0)^2 -\gamma v_\text{wind} x$.  We can easily see that the steady-state position found above is the minimum of this effective potential.

This example shows that the wind has the same effect as a linear contribution to the potential energy.  That statement agrees with the common intuition that moving downwind is like moving downhill, and moving upwind is like moving uphill.  More generally, the example also shows the procedure for calculating an effective potential:  First differentiate the Rayleigh dissipation function with respect to velocity, then set the velocity equal to zero, and then integrate with respect to the position.  In this article, we will apply that procedure to other generalized coordinates and generalized velocities in liquid-crystal physics.

Before going on, we should compare our analysis with the variational theory of Doi~\cite{Doi2011,Doi2013}.  Doi's theory also seeks to describe the behavior in terms of a single function.  That theory begins with the overdamped equation (\ref{overdamped}), and rewrites it in the form
\begin{equation}
0=-\frac{\partial R}{\partial\dot{x}},
\label{doi}
\end{equation}
where $R$ is a function that Doi calls the Rayleighian, defined by
\begin{equation}
R=\frac{\partial U}{\partial x}\dot{x}+D.
\end{equation}
Note that $R$ is different from $U_\text{eff}$, because $R$ is integrated with respect to $\dot{x}$ while $U_\text{eff}$ is integrated with respect to $x$.  For the simple example of a particle in the wind, we have
\begin{equation}
R=k(x-x_0)\dot{x}+\frac{1}{2}\gamma[\dot{x}-v_\text{wind}(x)]^2
\end{equation}
Equation~(\ref{doi}) then states that $R$ is minimized over velocity $\dot{x}$.  We emphasize that $R$ is \emph{not} minimized over position $x$.  Rather, minimization over $\dot{x}$ gives the equation of motion, and this equation must be solved to find the steady-state $x$.

As we understand it, Doi's theory is a useful way to derive equations of motion.  It is a variational theory in the sense that equations of motion are derived by differentiating $R$ with respect to velocity.  It is not actually a minimization theory, because the system does not move toward a minimum of $R$.  By contrast, the effective potential theory is a minimization theory in this stronger sense, because the system does move toward a minimum of $U_\text{eff}$.

\subsection{Two dimensions}

To see important limitations of the effective potential theory, consider 2D motion of a particle in the wind.  Suppose the particle position $\bm{r}(t)$, and the wind velocity field is $\bm{v}_\text{wind}(x,y)$, so that the Rayleigh dissipation function is $D=\frac{1}{2}\gamma|\dot{\bm{r}}-\bm{v}_\text{wind}(x,y)|^2$.  Following the same notation as above, the steady-equation of motion is
\begin{equation}
0=-\frac{\partial U}{\partial \bm{r}}-\left[\frac{\partial D}{\partial\dot{\bm{r}}}\right]_{\dot{\bm{r}}=0}
=-\frac{\partial U}{\partial \bm{r}}+\gamma\bm{v}_\text{wind}(x,y).
\end{equation}
We would like to rewrite that equation in the form
\begin{equation}
0=-\frac{\partial U_\text{eff}}{\partial \bm{r}},
\end{equation}
and hence we must define
\begin{align}
U_\text{eff}(x,y)&=U(x,y)+\int d\bm{r}\cdot\left[\frac{\partial D}{\partial\dot{\bm{r}}}\right]_{\dot{\bm{r}}=0}\nonumber\\
&=U(x,y)-\gamma\int d\bm{r}\cdot\bm{v}_\text{wind}(x,y).
\label{2dintegral}
\end{align}
We must now ask:  Does this integral have a single value, or does it depend on the integration path?  The answer depends on the curl of the wind velocity field.  If $\nabla\times\bm{v}_\text{wind}=0$, then the integral is independent of path, and hence the effective potential is a uniquely defined function.  By contrast, if $\nabla\times\bm{v}_\text{wind}\not=0$ (as in a hurricane), then the integral depends on the path, and the effective potential is not well defined.  This example demonstrates that the concept of an effective potential might or might not be useful in problems with more than one degree of freedom, depending on the generalized curl of the dissipative force.

For one more variation on this problem, suppose that the particle is constrained to move on a 1D curve in 2D, like a train constrained to move on a railroad track.  If the curve is not closed, then there is only a single integration path from one point to another in Eq.~(\ref{2dintegral}), and hence the effective potential is uniquely defined as in 1D.  However, if the curve is a closed loop, then the integration path might go around the loop once or multiple times.  Hence, the effective potential becomes a multi-valued function, with a discrete set of possible values.  We can still work with it, but we need to be careful with branch cuts, so that we only compare the effective potential of states on the same branch.  Several examples of this phenomenon in liquid-crystal physics will be given in the following sections.

\section{Uniform liquid crystal under simple shear flow}

For an example of an effective potential in liquid-crystal physics, consider a nematic phase under simple shear flow.  For simplicity, we work in 2D, so that the director is $\hat{\bm{n}}(t)=(\cos\theta(t),\sin\theta(t))$.  We assume the phase is uniform, so that the director may depend on time but not on position.

In hydrodynamic theory, there are two modes that dissipate energy:  the strain rate tensor, $A_{ij}=\frac{1}{2}(\partial_i v_j +\partial_j v_i)$, and the director totation with respect to the background fluid vorticity, $N_i=\dot{n}_i-\frac{1}{2}(\partial_j v_i -\partial_i v_j)n_j$.  The most general quadratic dissipation function density is then~\cite{Stewart2004}
\begin{align}
\label{dissipationgeneral}
D=&\frac{1}{2}\alpha_4 A_{ij}A_{ij}+\frac{1}{2}(\alpha_5+\alpha_6)n_i A_{ij}A_{jk}n_k \\
&+\frac{1}{2}\alpha_1 (n_i A_{ij}n_j)^2 +\frac{1}{2}\gamma_1 N_i N_i +\gamma_2 N_i A_{ij} n_j,\nonumber
\end{align}
where the $\alpha$ coefficients are the Leslie viscosities for fluid flow, $\gamma_1$ is the rotational viscosity for director rotation with respect to background fluid vorticity, and $\gamma_2$ is the torsion coefficient, representing a dissipative coupling between strain rate and director rotation.

For simple shear flow, we consider the fluid flow velocity profile is $\bm{v}=(v'y,0)$.  Hence, the dissipative modes become
\begin{equation}
A_{ij}=
\begin{pmatrix}
0            & \frac{v'}{2} \\
\frac{v'}{2} & 0
\end{pmatrix},
\quad
N_i=
\begin{pmatrix}
-\sin\theta \\
\cos\theta
\end{pmatrix}
\left[\dot{\theta}+\frac{v'}{2}\right],
\end{equation}
and the dissipation function becomes
\begin{align}
D=&\frac{1}{8}(\alpha_4+\alpha_5+\alpha_6)v'^2 +\frac{1}{8}\alpha_1 v'^2 \sin^2 2\theta \\
&+\frac{1}{2}\gamma_1 \left[\dot{\theta}+\frac{v'}{2}\right]^2 +\frac{1}{2}\gamma_2 v'\left[\dot{\theta}+\frac{v'}{2}\right]\cos2\theta.\nonumber
\end{align}

To convert this dissipation function into an effective potential acting on the steady-state angle $\theta$, we follow the procedure developed in Sec.~II(A).  First, we calculate the dissipative force acting on $\theta$ by differentiating the dissipation function with respect to $\dot{\theta}$,
\begin{equation}
-\frac{\partial D}{\partial\dot{\theta}}=\gamma_1 \left[\dot{\theta}+\frac{v'}{2}\right] +\frac{1}{2}\gamma_2 v'\cos2\theta.
\end{equation}
Next, we go to the steady-state case by setting $\dot{\theta}=0$, and obtain
\begin{equation}
-\left[\frac{\partial D}{\partial\dot{\theta}}\right]_{\dot{\theta}=0}=\frac{1}{2}\gamma_1 v'+\frac{1}{2}\gamma_2 v'\cos2\theta.
\end{equation}
Finally, we integrate with respect to $\theta$, and find the effective potential
\begin{equation}
U_\text{eff}=\int d\theta\left[\frac{\partial D}{\partial\dot{\theta}}\right]_{\dot{\theta}=0}
=\frac{1}{2}\gamma_1 v'\theta+\frac{1}{4}\gamma_2 v'\sin2\theta.
\label{dissipationpotential}
\end{equation}

\begin{figure}
\includegraphics[width=\columnwidth]{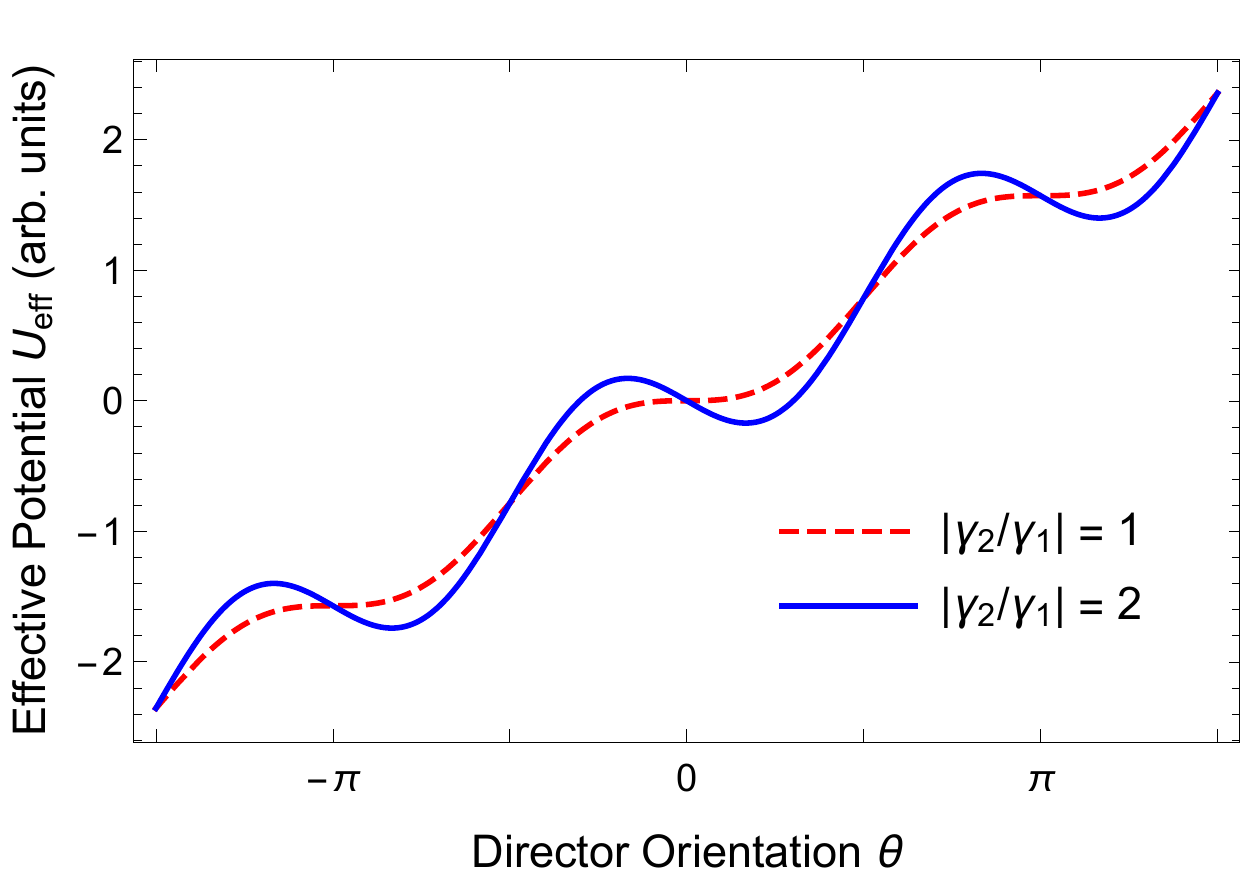}
\caption{(Color online) Effective potential $U_\text{eff}(\theta)$ for shear alignment of a uniform nematic liquid crystal in the 2D plane.  If the viscosity ratio $|\gamma_2/\gamma_1|\leq1$ (as in the dashed red curve), the effective potential has no local minima, and the director tumbles.  If $|\gamma_2/\gamma_1|>1$ (as in the solid blue curve), the effective potential has local minima, and the director aligns at one of those minima.  The plot is drawn with $\gamma_1>0$, $\gamma_2<0$ (as in typical experimental materials), and $v'>0$.}
\end{figure}

This expression for $U_\text{eff}(\theta)$ is plotted in Fig.~1.  From this figure, we can see that the effective potential can have two possible shapes, depending on the viscosity ratio $|\gamma_2/\gamma_1|$.  If $|\gamma_2/\gamma_1|\leq1$, the effective potential is a monotonically increasing or decreasing function of $\theta$, with no local minima.  In that case, the director tumbles continuously toward lower values of $U_\text{eff}$.  The direction of tumbling depends on the sign of $v'$.  A positive shear flow $v'>0$ induces tumbling with $\dot{\theta}<0$, while $v'<0$ induces $\dot{\theta}>0$.  By contrast, if $|\gamma_2/\gamma_1|>1$, then the the effective potential has a series of local minima, and the director aligns at one of these minima.  To find the alignment orientation, we can solve $\partial U_\text{eff}/\partial\theta=0$, which gives
\begin{equation}
\theta=\frac{1}{2}\cos^{-1}\left[-\frac{\gamma_1}{\gamma_2}\right].
\label{leslieangle}
\end{equation}

Of course, the theory of shear alignment was developed and the alignment angle was calculated many years ago.  This calculation is normally expressed as a balance of stresses, rather than as the minimization of any function.  We suggest that the effective potential provides several insights into this classic calculation.  First, we can see that the magnitude of shear flow $v'$ does not affect the alignment angle, but it does determine the overall magnitude of $U_\text{eff}$.  Hence, a stronger shear flow gives a stronger alignment at the same angle.  Second, we can see that the arccosine in Eq.~(\ref{leslieangle}) has multiple values, and some of them are local minima of $U_\text{eff}$ while others are local maxima.  Reversing the sign of $v'$ reverses the sign of $U_\text{eff}$, and hences exchanges the minima and maxima.  Third, we can see how alignment by shear flow competes with other alignment mechanisms, such as alignment by an applied magnetic field.  The effective potential $U_\text{eff}$ derived from shear flow can simply be added to the magnetic free energy $-(\Delta\chi/2\mu_0)(\bm{H}\cdot\hat{\bm{n}})^2$, or to any other free energy terms, and the equilibrium director can be determined by minimization of the total effective free energy.

We note that the same physical state of the liquid crystal can be described by multiple angles ($\theta$, $\theta\pm\pi$, $\theta\pm2\pi$, \ldots), and hence the same physical state has multiple values of the effective potential.  This problem of a multi-valued effective potential is analogous to the problem of motion in a hurricane on a circular railroad track, as discussed in Sec.~II(B).  We can interpret the analogy by working in terms of the nematic director $\hat{\bm{n}}=(\cos\theta,\sin\theta)$.  The two components of $\hat{\bm{n}}$ are analogous to the 2D plane, and the constraint $|\hat{\bm{n}}|=1$ is analogous to the track that constrains the motion.  Alternatively, we can interpret the analogy by working in terms of the 2D nematic order tensor $Q_{ij}=S(2n_i n_j-\delta_{ij})$, where $S$ is the scalar order parameter.  Here, the two independent components $Q_{xx}=-Q_{yy}=S\cos2\theta$ and $Q_{xy}=Q_{yx}=S\sin2\theta$ are analogous to the 2D plane, and the thermal free energy that determines $S$ provides an approximate constraint on the motion, analogous to the track.  In either case, the effective potential is locally well-defined, under small rotations of the physical state.  However, there are different branches of the effective potential, depending on which branch of the angle $\theta$ is chosen, and it is not meaningful to compare different branches.

As we mentioned in the Introduction, previous theoretical research has studied the effects of shear flow on the isotropic-nematic transition~\cite{Olmsted1990,Olmsted1992}.  Near this transition, one must consider the two independent components of the nematic order tensor \emph{without} a strong constraint on the scalar order parameter $S$.  This problem is analogous to motion in the 2D plane without a track.  In that case, the effective potential involves an integral that is path-dependent, and hence is not even locally well-defined.  Thus, the concept of an effective potential may not be useful for that problem.

\section{Reverse tilt domains under simple shear flow}

In this section, we apply the concept of minimizing an effective potential to a nonuniform liquid crystal.  For a simple example of a nonuniform liquid crystal with no defects, we consider the dynamics of reverse tilt domains under shear flow.

\begin{figure}
\includegraphics[width=\columnwidth]{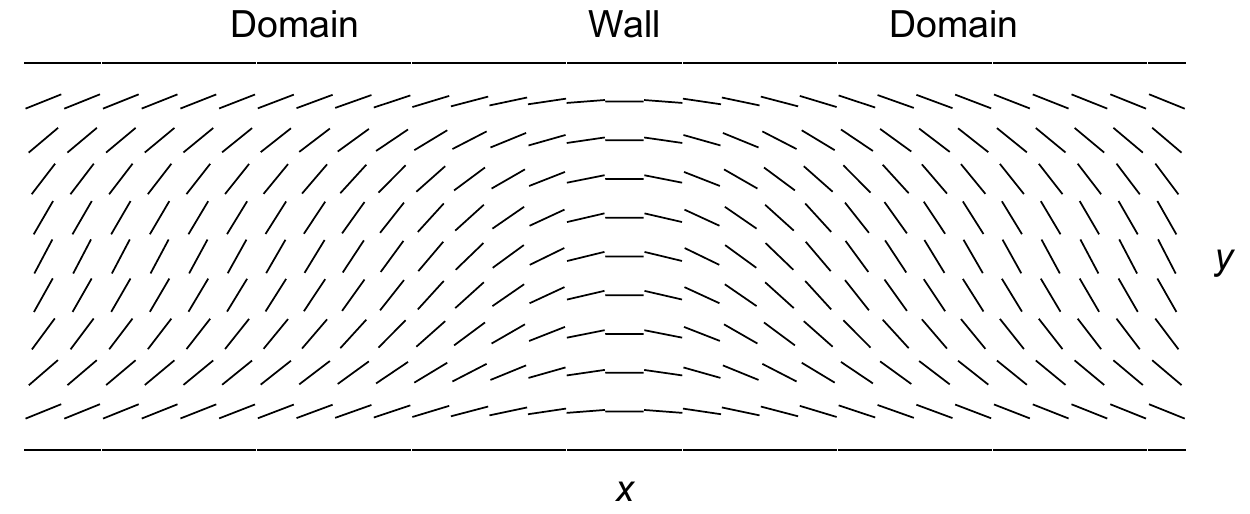}
\caption{Example of reverse tilt domains separated by a domain wall.  The top and bottom surfaces of the cell have strong planar anchoring, and an electric or magnetic field is applied across the thickness of the cell, in the $y$ direction.}
\end{figure}

The concept of reverse tilt domains is illustrated in Fig.~2.  When a nematic liquid crystal is confined to a narrow cell with strong anchoring conditions, and then subjected to an electric or magnetic field above the Fr\'eedericksz transition, the director field becomes nonuniform across the thickness of the cell.  It aligns with the anchoring direction near the walls, and tilts toward the field direction in the interior.  Ideally, it would tilt in the same orientation for all horizontal positions.  However, in typical experiments, the director tilts in a positive or negative orientation in different parts of the cell.  The regions of different tilt are called reverse tilt domains, and the narrow regions between these domains are domain walls~\cite{Shenoy1999,Lee2006,Wang2007}.

In the simple geometry of Fig.~2, there is a symmetry between the domains of positive and negative tilt on the left and right sides.  Because of that symmetry, the domains have equal free energy, and neither domain will grow or shrink.  In some cases, an experimenter might want to break the symmetry, in order to favor positive or negative tilt.  One method to break the symmetry is to prepare surfaces with a pretilt.  With modified anchoring conditions, the surface can induce the director field to have a slight tilt away from the planar orientation, in a positive or negative direction.  Another method is to tilt the applied electric or magnetic field.  This tilted field favors a the corresponding tilted domain, and disfavors the other.  These two methods for breaking the symmetry can certainly be described by a free energy.  With a specified surface pretilt or bulk tilted field, one can calculate the free energy difference between domains of positive and negative tilt.  Based on this free energy difference, one can predict how one domain grows and the other shrinks.

Now consider a third method to break the symmetry by applying a shear flow $\bm{v}=(v' y,0)$.  Under shear flow, one type of domain is closer to the favored shear alignment angle, and hence is compatible with the flow.  By comparison, the other type of domain is farther from the favored angle, and is less compatible with the flow.  The question is:  Can this method for breaking the symmetry be described by an effective potential?  Under shear flow, is there an effective potential difference between domains of positive and negative tilt, which determines how one domain grows and the other shrinks?

To answer that question, we extend the effective potential argument of the previous section to a nonuniform liquid crystal in a reverse tilt domain.  Suppose the director field is $\hat{\bm{n}}(y,t)=(\cos\theta(y,t),\sin\theta(y,t))$, and the applied electric field is $\bm{E}=(0,E)$.  The free energy density then becomes
\begin{equation}
F=\frac{1}{2}K\left(\frac{\partial\theta}{\partial y}\right)^2 - \frac{1}{2}\epsilon_0 \Delta\epsilon E^2 \sin^2 \theta,
\end{equation}
where $\Delta\epsilon$ is the dielectric anisotropy, and we assume a single Frank constant $K$.  Following the argument of Eqs.~(\ref{dissipationgeneral})--(\ref{dissipationpotential}), we construct the dissipation function, differentiate it with respect to $\dot{\theta}$ to find the dissipative force, and integrate it with respect to $\theta$ to find the dissipative part of the effective potential density,
\begin{equation}
\int d\theta\left[\frac{\delta D}{\delta\dot{\theta}}\right]_{\dot{\theta}=0}
=\frac{1}{2}\gamma_1 v'\theta+\frac{1}{4}\gamma_2 v'\sin2\theta.
\end{equation}
Combining the free energy and the dissipative terms gives the total effective potential density
\begin{align}
U_\text{eff}=&\frac{1}{2}K\left(\frac{\partial\theta}{\partial y}\right)^2 - \frac{1}{2}\epsilon_0 \Delta\epsilon E^2 \sin^2 \theta \nonumber\\
&+\frac{1}{2}\gamma_1 v'\theta+\frac{1}{4}\gamma_2 v'\sin2\theta.
\label{effectivepotentialRTD}
\end{align}

Now that we have derived the effective potential density, we can use it to calculate the steady-state director field, just as we are accustomed to using the free energy density to calculate the equilibrium director field.  In particular, we can construct the Euler-Lagrange equation
\begin{align}
0=&\frac{\delta U_\text{eff}}{\delta\theta}\\
=&-K\frac{\partial^2 \theta}{\partial y^2}-\frac{1}{2}\epsilon_0 \Delta\epsilon E^2 \sin2\theta+\frac{1}{2}\gamma_1 v'+\frac{1}{2}\gamma_2 v'\cos2\theta. \nonumber
\end{align}
This equation is equivalent to the Ericksen-Leslie equation for the steady-state director field in the Fr\'eedericksz transition under shear flow, and it is difficult to solve exactly.

As a simpler alternative, we make a variational ansatz for the steady-state director field $\theta(y)=\theta_0 \cos(\pi y/d)$, which satisfies the planar anchoring conditions at $y=\pm d/2$.  We insert this ansatz into the effective potential density~(\ref{effectivepotentialRTD}) and then average over the thickness of the cell to obtain
\begin{align}
U_\text{eff}^\text{average}=&\frac{\pi^2 K \theta_0^2}{4 d^2}+\frac{\epsilon_0 \Delta\epsilon E^2}{2}\left[J_0 (2\theta_0)-1\right]\nonumber\\
&+\frac{\gamma_1 v' \theta_0}{\pi}+\frac{\gamma_2 v'}{4} \bm{H}_0 (2\theta_0),
\end{align}
where $J_0$ and $\bm{H}_0$ are the Bessel and Struve functions, respectively.  Near the Fr\'eedericksz transition, for $\theta_0 \ll 1$, we can expand as a power series in $\theta_0$ to obtain
\begin{align}
U_\text{eff}^\text{average}=&\frac{(\gamma_1 + \gamma_2)v' \theta_0}{\pi}
+\frac{\epsilon_0 \Delta\epsilon(E_c^2-E^2)\theta_0^2}{2}\nonumber\\
&-\frac{4\gamma_2 v' \theta_0^3}{9\pi}
+\frac{\epsilon_0 \Delta\epsilon \theta_0^4}{8}+\cdots,
\end{align}
where $E_c=[(\pi^2 K)/(2\epsilon_0 \Delta\epsilon d^2)]^{1/2}$ is the critical field.  From this series, we can see how shear flow changes the Fr\'eedericksz transition.  In the absence of flow, for $v'=0$, the effective potential has an exact symmetry between positive and negative tilt $\theta_0$.  Above the critical field, there are two minima at $\theta_0 = \pm[2(1-E_c^2/E^2)]^{1/2}$, and these two minima have the same effective potential.  By contrast, for $v'\neq0$, the power series has odd terms that break the symmetry between positive and negative tilt.  The shear flow acts as an effective field that favors one sign of $\theta_0$.  Hence, the two minima are shifted, and one minimum becomes lower in effective potential than the other.  For that reason, the domain wall between neighboring domains in Fig.~2 will move, so that the domain with lower effective potential will grow, and the domain with higher effective potential will shrink.

We have performed numerical simulations of the dynamics of two reverse tilt domains separated by a wall, using the same $Q$ tensor method as in our previous article~\cite{Tang2019}.  These simulations confirm that the the domain of lower effective potential grows, and the domain of higher effective potential shrinks.  The velocity of the wall between these domains is proportional to the shear rate $v'$, and hence to the difference of effective potential between the two domains.  These results confirm that the effective potential provides a useful way to understand which domain is favored by the imposed shear flow.

\section{Dowser and bowser states under Poiseuille flow}

In this section, we consider the motion of a disclination between the dowser and bowser states in a narrow liquid-crystal cell.  This problem has already been studied using an effective potential concept by Emer{\v{s}}i{\v{c}} \emph{et al}.~\cite{Emersic2019}.  We show that their effective potential concept is consistent with the general approach presented in this article.

The dowser state has been found experimentally in several studies by Pieranski \emph{et al.}~\cite{Pieranski2016a,Pieranski2016b,Pieranski2017,Pieranski2019a,Pieranski2019b}.  It has the structure shown schematically in Fig.~3a.  Suppose that a liquid-crystal cell has strong homeotropic anchoring on both sides.  The simplest director configuration is just a uniform vertical alignment, shown in the right side of the figure.  However, under some circumstances, the director field might form a more complex state, shown in the left side of the figure, which Pieranski \emph{et al.}\ have called the dowser state.  In the dowser state, the director rotates through $180^\circ$ from the top to the bottom surface.  At any interface where the uniform and dowser states meet, the liquid crystal must have a disclination of topological charge $\pm1/2$, shown in the middle of the figure.

\begin{figure}
\includegraphics[width=\columnwidth]{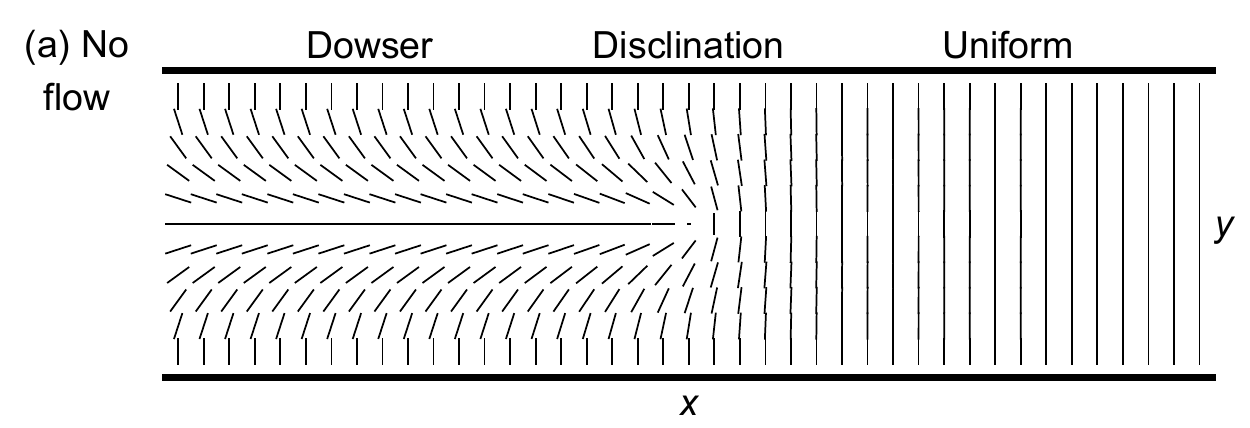}
\includegraphics[width=\columnwidth]{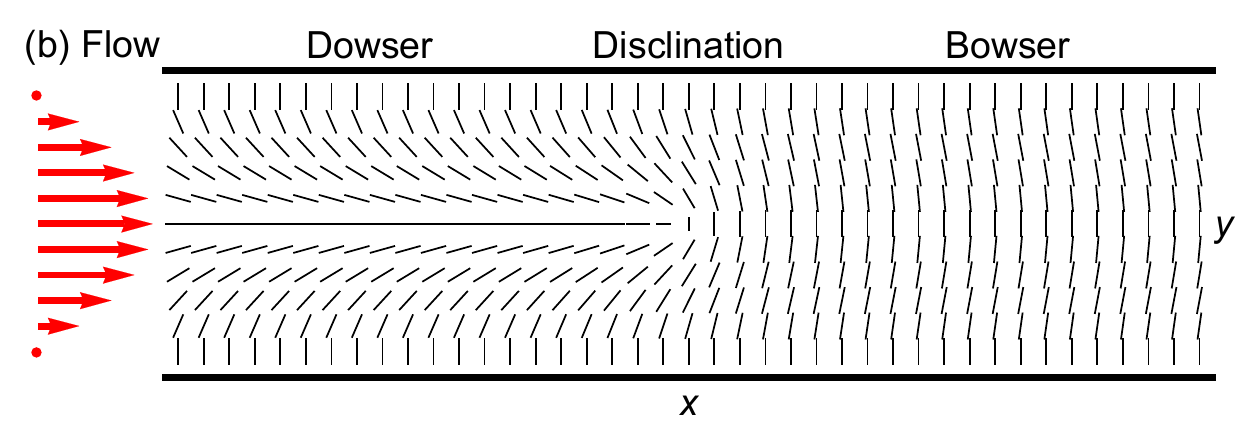}
\caption{(Color online) (a)~Dowser and uniform states, separated by a disclination, in the absence of flow.  (b)~Dowser and bowser states, separated by a disclination, in the presence of Poiseuille flow (indicated by the red arrows).}
\end{figure}

In general, the dowser has a higher elastic free energy than the uniform state, because the dowser has director gradients while the uniform state does not.  In equilibrium, the dowser region shrinks and the uniform region grows, so that the liquid crystal can reduce its total free energy.  This process occurs by motion of the disclination at the interface.  In the example of Fig.~3a, the disclination moves to the left to reduce the dowser region and increase the uniform region.

In the experiments of Ref.~\cite{Emersic2019}, Emer{\v{s}}i{\v{c}} \emph{et al}.\ stabilize the dowser state by applying planar Poiseuille flow.  Poiseuille flow is a parabolic flow profile $\bm{v}=({v_\text{max}(1-4y^2/d^2)},0)$, so that the velocity is zero at the top and bottom surfaces $y=\pm d/2$, and is maximum in the middle of the cell, as shown in red in Fig.~3b.  Poiseuille flow includes shear flow $\partial v_x/\partial y$ with one sign in the upper half of the cell, and the opposite sign in the lower half of the cell.  This shear flow induces alignment of the director field with one sign in the upper half and the opposite sign in the lower half.  The profile of shear alignment angle across the thickness is similar to the director profile in the dowser state.  Hence, one might expect the dowser state to be compatible with Poiseuille flow.  By contrast, when the uniform state is exposed to Poiseuille flow, it deforms into the bow-like director profile on the right of Fig.~3b, which Emer{\v{s}}i{\v{c}} \emph{et al}.\ call the ``bowser'' state.  One might expect the bowser state to be less compatible than the dowser with Poiseuille flow.  Indeed, the experiments demonstrate that a large enough Poiseuille flow causes the dowser state to grow and the bowser state to shrink.

We would like to understand the stabilization of the dowser state through the same type of effective potential concept as in the previous sections.  In particular, we would like to see how the Poiseuille flow affects the effective potential of the dowser in comparison with the bowser state.

For this calculation, we repeat the argument of Sec.~IV with two small differences:  There is no applied electric field $E=0$, and the shear rate for Poiseuille flow is given by $v' = \partial v_x /\partial y = -8v_\text{max}y/d^2$.  Hence, the effective potential density becomes
\begin{align}
U_\text{eff}=&\frac{K}{2}\left(\frac{\partial\theta}{\partial y}\right)^2 -\frac{4\gamma_1 v_\text{max}y\,\theta}{d^2}-\frac{2\gamma_2 v_\text{max}y\sin2\theta}{d^2}.
\label{effectivepotentialbowserdowser}
\end{align}
We must integrate this density over the thickness of the cell for both bowser and dowser states.

For the bowser state, the simplest assumption for the director field is just the uniform $\theta(y)=\pi/2$.  By putting this assumption into the effective potential~(\ref{effectivepotentialbowserdowser}) and averaging over the thickness of the cell, we obtain just 
\begin{equation}
U_\text{eff}^\text{bowser}=0.
\label{ueffbowser}
\end{equation}

For a more detailed model of the bowser, we could use the higher-order expression $\theta(y)=(\pi/2)+\theta_0 \sin(2\pi y/d)$, which is vertical at the top and bottom surfaces as well as in the center of the cell, and is tilted away from vertical in opposite senses in the upper and lower halves.  Putting this assumption into the effective potential, expanding as a power series for small $\theta_0$, and averaging over the cell thickness gives
\begin{equation}
U_\text{eff}^\text{average}=-\frac{2(\gamma_1-\gamma_2)v_\text{max}\theta_0}{\pi d}+\frac{\pi^2 K\theta_0^2}{d^2}
\end{equation}
Minimizing this expression over $\theta_0$ gives 
\begin{equation}
\theta_0 = \frac{(\gamma_1-\gamma_2)v_\text{max}d}{\pi^3 K}.
\end{equation}
Hence, Poiseuille flow transforms the uniform state into a bowser with director variation $\theta_0$ proportional to $v_\text{max}$.  Putting that expression back into the effective potential gives corrections to Eq.~(\ref{ueffbowser}).  However, we will not need those corrections in the argument below.

For the dowser state, we must be careful to choose the appropriate quadrant for the angle $\theta$.  In choosing the quadrant, we use the following physical argument:  Suppose that the disclination moves to the right, so that the bowser is transformed into the dowser.  In the upper half of the cell, the director rotates counter-clockwise to $\theta>\pi/2$.  In the bottom half of the cell, the director rotates clockwise to $\theta<\pi/2$.  Hence, the simplest assumption for the director field is
\begin{equation}
\theta(y)=
\begin{cases}
\pi(1-y/d), & \text{for } 0<y<d/2,\\
-\pi y/d, & \text{for } -d/2<y<0.
\end{cases}
\label{dowseransatz}
\end{equation}
In other words, we must put a branch cut for $\theta$ at $y=0$, the same height as the disclination.  This choice of quadrant corresponds to the issue discussed in Sec.~II(B) for the effective potential of a particle  that is constrained to move on a closed loop.  The effective potential is a multi-valued function, and our choice of the branch must be consistent to compare the bowser and dowser states.

By putting assumption~(\ref{dowseransatz}) for the director field into the effective potential~(\ref{effectivepotentialbowserdowser}) and averaging over the cell thickness, we obtain
\begin{equation}
U_\text{eff}^\text{dowser}=\frac{\pi^2 K}{2d^2}-\frac{v_\text{max}}{d}\left(\frac{\pi\gamma_1}{6}-\frac{\gamma_2}{\pi}\right).
\label{ueffdowser}
\end{equation}
In typical experimental materials, we have $\gamma_2<0$, and hence the term in parentheses is positive.

We can now compare the effective potentials of the bowser and the dowser in Eqs.~(\ref{ueffbowser}) and~(\ref{ueffdowser}), respectively.  When the Poiseuille flow velocity $v_\text{max}$ is low, the unfavorable elastic term dominates the effective potential for the dowser, and hence $U_\text{eff}^\text{bowser}<U_\text{eff}^\text{dowser}$.  By contrast, when $v_\text{max}$ is high, the favorable dissipative term dominates the effective potential for the dowser, and hence $U_\text{eff}^\text{bowser}>U_\text{eff}^\text{dowser}$.  The two effective potentials are equal at the velocity
\begin{equation}
v_\text{max}^*
=\frac{\pi^2 K}{2d}\left(\frac{\pi\gamma_1}{6}-\frac{\gamma_2}{\pi}\right)^{-1}.
\end{equation}
Hence, the effective potential concept shows quantitatively that the bowser is preferred for $v_\text{max}<v_\text{max}^*$ and the dowser for $v_\text{max}>v_\text{max}^*$.

We can extend this concept to model the motion of the disclination driven by a difference of effective potential.  If the disclination moves to the right by a distance $\delta x$, then an area $d\,\delta x$ is transformed from bowser to dowser, so the total effective potential changes by $d\,\delta x(U_\text{eff}^\text{dowser}-U_\text{eff}^\text{bowser})$.  Hence, the effective potential generates a force of $F_\text{potential}=-d(U_\text{eff}^\text{dowser}-U_\text{eff}^\text{bowser})$ acting on the disclination.  In addition, the motion of the disclination with velocity $u$, with respect to the Poiseuille flow with velocity $v_\text{max}$, generates a drag force of $F_\text{drag}=-\eta(u-v_\text{max})$, where $\eta$ is the drag coefficient for the disclination.  A classic result for this drag coefficient~\cite{Imura1973}, discussed in our previous paper~\cite{Tang2019}, is $\eta=(\pi\gamma_1/4)\log[d/(2r_\text{core})]$, where $r_\text{core}$ is the disclination core radius.  (This minimal model assumes equal Frank constants, flow viscosity $\alpha_4$ much greater than rotational viscosity $\gamma_1$, and all other viscosities equal to zero.)  When the drag force cancels the effective potential force, the disclination moves at the steady-state velocity
\begin{align}
\label{dowservelocityplanar}
u&=v_\text{max}-\frac{d}{\eta}(U_\text{eff}^\text{dowser}-U_\text{eff}^\text{bowser})\\
&=v_\text{max}+\frac{2v_\text{max}[1-(6\gamma_2)/(\pi^2\gamma_1)]-(6\pi K)/(d\gamma_1)}{3\log[d/(2r_\text{core})]}.\nonumber
\end{align}
In this result, the first term shows that the disclination is carried along by the Poiseuille flow, and the second term shows the extra (positive or negative) motion induced by the difference of effective potential between bowser and dowser.

\begin{figure}
\includegraphics[width=\columnwidth]{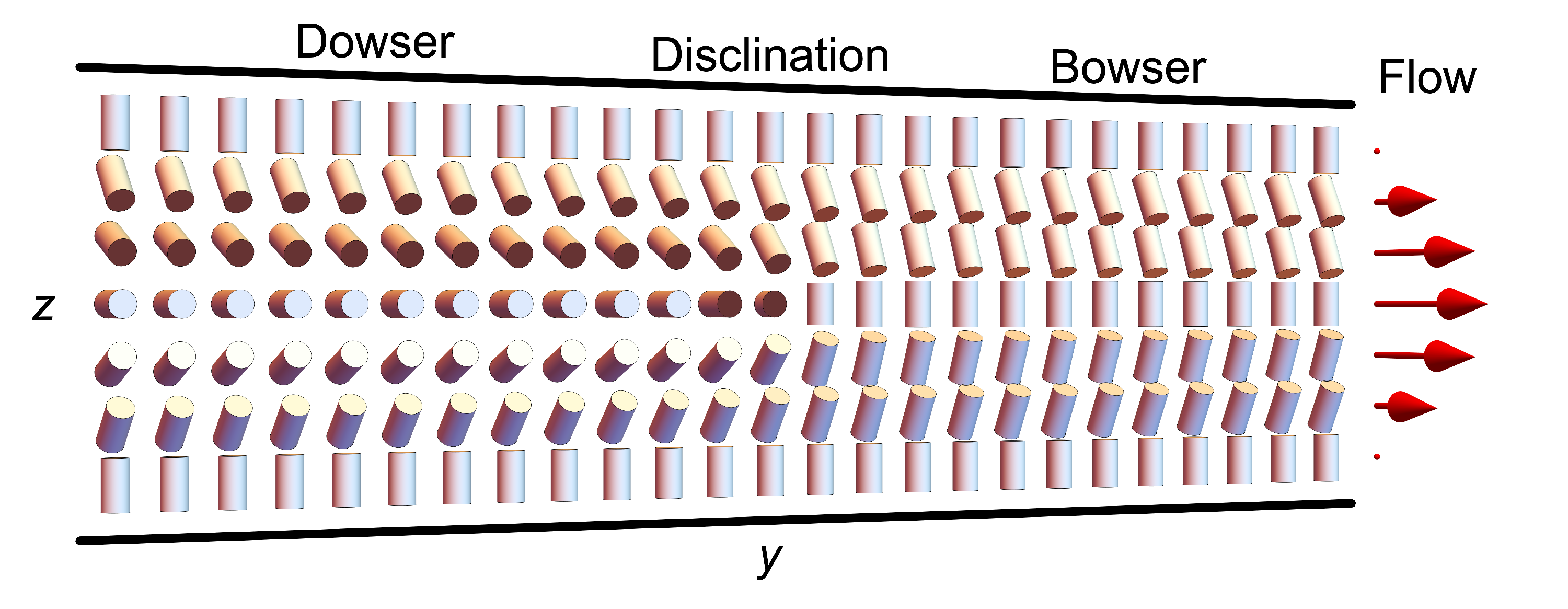}
\caption{(Color online) Three-dimensional (3D) geometry of dowser and bowser states, separated by a twisted disclination, in the presence of Poiseuille flow (indicated by the red arrows).}
\end{figure}

The 2D geometry of a disclination between bowser and dowser necessarily shows both motion carried along by the Poiseuille flow and motion induced by the effective potential difference.  As a conceptual exercise, one might want to separate these two effects, and see only motion induced by the effective potential difference.  For that reason, we consider the 3D system shown in Fig.~4.  In this geometry, the Poiseuille flow is in the $x$ direction, the director field is in the $(x,z)$ plane, and the boundary between dowser and bowser occurs at $y=0$.  Hence, the disclination runs along the $x$-axis, and the director around the disclination has a 3D twisted structure.  Hence, the Poiseuille flow carries the disclination along its own length, in the $x$ direction.  By contrast, the effective potential difference pushes the disclination in the $y$ direction, so that the bowser grows and the dowser shrinks, or vice versa.  The force from the effective potential is the same as in the previous case, while the drag force is $F_\text{drag}=-\eta u$.  Hence, the forces balance when the disclination moves at the steady-state velocity
\begin{equation}
u=\frac{2v_\text{max}[1-(6\gamma_2)/(\pi^2\gamma_1)]-(6\pi K)/(d\gamma_1)}{3\log[d/(2r_\text{core})]}
\label{dowservelocitytwisted}
\end{equation}
in the $y$ direction.

To test the effective potential approach, we perform numerical simulations of a moving disclination between bowser and dowser states, under imposed Poiseuille flow.  We use the same $Q$ tensor method as in our previous article~\cite{Tang2019}.  In this method, the free energy density is
\begin{equation}
F=-\frac{1}{4}a Q_{ij}Q_{ij} + \frac{1}{16} b (Q_{ij}Q_{ij})^2 + \frac{1}{16}L(\partial_k Q_{ij})(\partial_k Q_{ij}),
\end{equation}
and the dissipation function density is
\begin{equation}
D=\frac{1}{16}\Gamma_1 B_{ij}B_{ij} + \frac{1}{4}\Gamma_2 B_{ij}A_{ij} + \frac{1}{2}\alpha_4 A_{ij}A_{ij},
\end{equation}
where
\begin{align}
B_{ij}&=\dot{Q}_{ij} - \omega_m (\epsilon_{mlj}Q_{il}+\epsilon_{mli}Q_{lj}),\nonumber\\
A_{ij}&=\frac{1}{2}(\partial_i v_j + \partial_j v_i),\nonumber\\
\bm{\omega}&=\bm{\nabla}\times\bm{v},
\end{align}
and $\epsilon_{ijk}$ is the Levi-Civita symbol.  The coefficients in this tensor representation are related to the coefficients in the director representation by $K = L S^2$, $\gamma_1 = \Gamma_1 S^2$, and $\gamma_2 = \Gamma_2 S$, where $S$ is the scalar order parameter.  For the simulations, we use parameters $d=2$ $\mu$m, $K=10$ pN, $\gamma_1 = 0.08$ Pa s, $\gamma_2 = -0.09$ Pa s, similar to the liquid crystal 5CB.  We choose $a$ and $b$ so that $S_\text{bulk}=(a/b)^{1/2}=1$ and $r_\text{core}=(K/a)^{1/2}=0.2$ $\mu$m.  (The core radius $r_\text{core}$ must be exaggerated for the numerical algorithm, but the results are not very sensitive to this value.)  We impose the Poiseuille flow profile, and then solve the hydrodynamic equation for the director field to find the steady-state velocity of the disclination.

\begin{figure}
\includegraphics[width=\columnwidth]{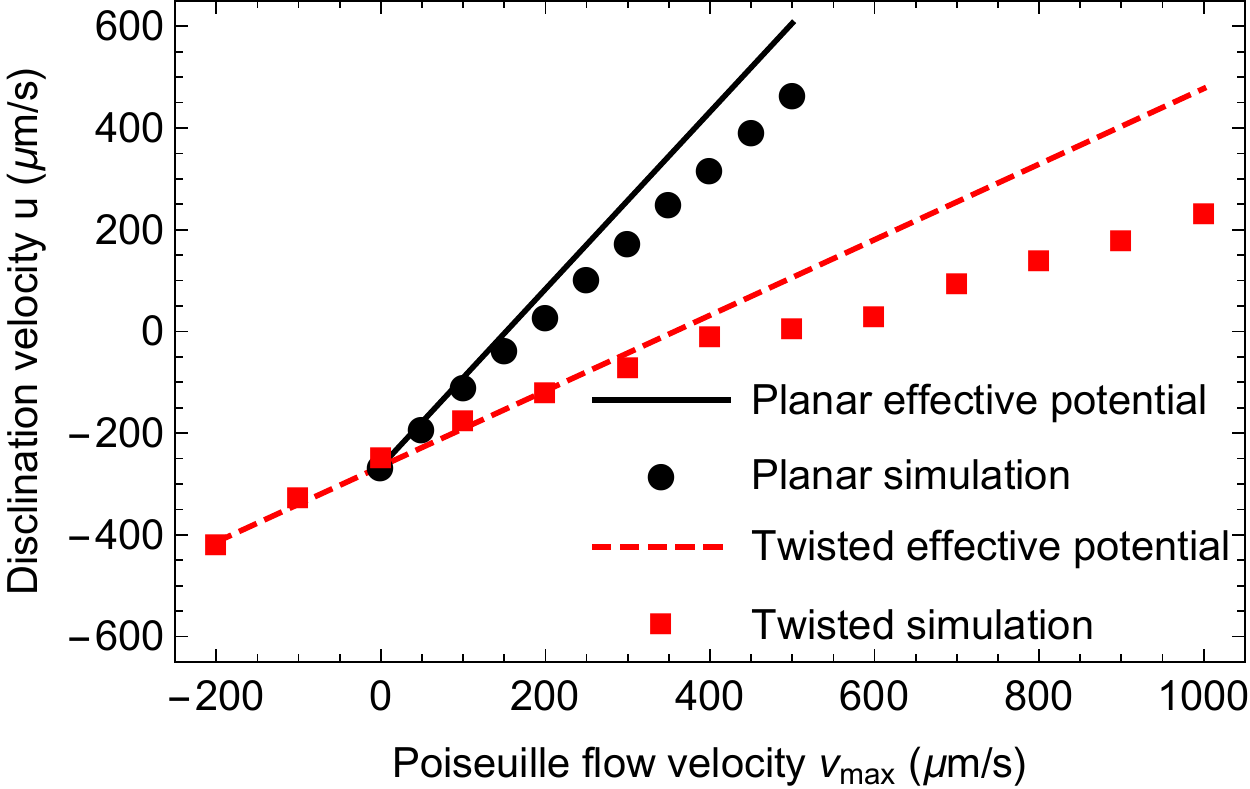}
\caption{(Color online) Simulation results for the disclination velocity $u$ as a function of Poiseuille flow velocity $v_\text{max}$, in comparison with the predictions of the effective potential theory, which are in Eq.~(\ref{dowservelocityplanar}) for the planar disclination and Eq.~(\ref{dowservelocitytwisted}) for the twisted disclination.  Numerical parameters are given in the text.}
\end{figure}

In Fig.~5, the black circles show the simulation results for the disclination velocity $u$ as a function of the Poiseuille flow velocity $v_\text{max}$.  By comparison, the black solid line shows the effective potential prediction of Eq.~(\ref{dowservelocityplanar}).  These two calculations show consistent behavior.  When $v_\text{max}=0$, the disclination moves to the left, with $u<0$.  The dowser shrinks and the bowser grows, so that the system can reduce its elastic free energy.  By comparison, when $v_\text{max}$ becomes large enough in the positive direction, the dowser is stabilized, and then the disclination moves to the right, with $u>0$, so that the dowser grows and the bowser shrinks.  We can regard this change as driven by the dissipative part of the effective potential.  Although the simulations and theory agree very well for small $v_\text{max}$, there are some discrepancies for larger $v_\text{max}$.  To address these discrepancies, we can add corrections to the effective potential calculation by allowing the director field to respond to the Poiseuille flow, but we do not present those results here.

We also perform numerical simulations of the twisted disclination between bowser and dowser.  These simulations use the same procedure as for the planar disclination, except that the Poiseuille flow is in the $x$ direction and the $Q$ tensor is in the $(x,z)$ plane.  The simulation results are shown by the red squares in Fig.~5, while the effective potential prediction of Eq.~(\ref{dowservelocitytwisted}) is shown by the red dashed line.  Again, the simulations and theory agree very well for small $v_\text{max}$, although there are some discrepancies for larger $v_\text{max}$.  We can see that the slope of $u$ as a function of $v_\text{max}$ is smaller for the twisted disclination than for the planar disclination, because the twisted disclination moves in response to effective potential differences but is not carried by Poiseuille flow.

Our effective potential theory for the motion of the disclination between bowser and dowser is equivalent to the theory developed by Emer{\v{s}}i{\v{c}} \emph{et al}.~\cite{Emersic2019}.  The significance of our work is to show how this theory fits into the general formalism for an effective potential that includes dissipative contributions.  In particular, we can see that this concept is the same effective potential that enters into shear alignment at the Leslie angle, and into the motion of reverse tilt domains.

\section{Discussion}

From the classical mechanics analogy in Sec.~II, we can see that the effective potential concept does not always apply.  The main issue that determines whether it works is the number of degrees of freedom.  If there is only one degree of freedom, the effective potential can be defined unambiguously.  If there is more than one degree of freedom, the effective potential might or might not be a uniquely defined quantity, depending on whether the dissipative force has a nonzero curl.  In an intermediate case, a system might have a closed path within a space with more than one degree of freedom, like a train on a circular track.  In that case, the effective potential is a multi-valued function, which still might be useful provided that one is careful with branches of the function.

A 2D nematic liquid crystal has both a magnitude $S$ and a direction $\theta$ of orientational order.  For that reason, it really is a system with more than degree of freedom.  However, in many cases, the magnitude is approximately fixed, and only the direction can vary.  In that sense, it is analogous to a train on a circular track of allowed states.  Hence, the effective potential is a multi-valued function, which depends on which branch of the angle $\theta$ is chosen.  Indeed, the multi-valued nature can be seen because the effective potential depends on $\theta$, not just on $\sin2\theta$ and $\cos2\theta$.

This multi-valued nature of the effective potential might or might not be important, depending on the problem.  For shear alignment in a uniform system, the multi-valued nature is not important as long as the director remains near a specific angle.  However, it becomes important if the director can tumble through a full circle.  For the reverse tilt domain problem, the surface anchoring constrains the director near a specific angle.  Hence, all physical states are on a single branch of the $\theta$ function, and the multi-valued nature of the effective potential is not important.  For the dowser and bowser, there is a disclination at the interface between the two states, and hence we cannot work consistently with a single branch of the $\theta$ function.  Even so, we can still compare the effective potentials of dowser and bowser, provided that we choose the quadrant of angle in a consistent way that matches the physical motion of the disclination. 

In conclusion, this article has shown that the effective potential concept is a useful way to think about shear alignment of liquid crystals.  Using the effective potential, we can see that one state is more favorable than another in the presence of imposed shear flow.  Hence, the effects of shear flow can be understood as minimization of effective potential, just as the effects of applied fields or surface alignment can be understood as minimization of the free energy.  This type of argument is more intuitive than just solving the hydrodynamic equations, and we expect that it will be useful for a range of nonequilibrium alignment problems.

\acknowledgments

We would like to thank S. \v{C}opar, P.~D. Olmsted, and U. Tkalec for helpful discussions.  This work was supported by National Science Foundation Grant DMR-1409658.

\bibliography{effectivepotential2}

\end{document}